\documentclass[aps,prl,twocolumn,showpacs,superscriptaddress,groupedaddress]{revtex4-1}
\usepackage[pdftex]{color}
\usepackage[utf8]{inputenc}
\usepackage[english]{babel}
\usepackage{amsfonts}
\usepackage{amssymb}

\usepackage{amsmath}
\usepackage{graphicx}
\usepackage[caption=false]{subfig}

\newcommand{\Ket}[1]{\left|#1\right>}

\newcommand{\BraKet}[2]{\left<#1|#2\right>}
\newcommand{\KetBra}[2]{|#1\rangle\langle#2|}

\newcommand{\expo}[1]{\mathrm{exp}\left(#1\right)}
\newcommand{\conm}[2]{\left[#1,#2\right]}

\begin{document}

\title{Controlling open quantum systems using fast transitions}
\author{Pablo M. Poggi}
\author{Fernando C. Lombardo}
\author{Diego A. Wisniacki}
\affiliation{Departamento de F\'isica Juan Jos\'e Giambiagi, 
FCEyN UBA, Facultad de Ciencias Exactas y Naturales, 
Ciudad Universitaria, Pabell\'on I, 1428 Buenos Aires, Argentina - IFIBA}

\pacs{03.67.Lx, 03.65.Yz}

\begin{abstract}
Unitary control and decoherence appear to be irreconcilable in quantum mechanics. When a quantum system interacts with an environment, control strategies usually fail due to decoherence. In this article we implement a time-optimal unitary control protocol suitable for quantum open systems. The method is based on succesive diabatic and sudden switch transitions in the avoided crossings of the energy spectra of closed systems. We show that the speed of this control protocol meets the fundamental bounds imposed by the quantum speed limit, thus making this scheme ideal for application where decoherence needs to be avoided. We show that this method can achieve complex control strategies with high accuracy in quantum open systems.
\end{abstract}

\maketitle

\textit{Introduction} -  Quantum control is a fundamental goal in different areas, including physical chemistry, nanoscience and quantum information processing \cite{bib:chem,bib:nano,bib:qc}. In fact, as the manipulation of quantum systems will be the basis for future technological applications, the development of different control strategies is a task of major interest nowadays. 

An unitary control scheme consists of engineering the time dependence of one or several external parameters (i.e. electric fields) which manipulate the temporal evolution of a quantum system in order to obtain a desired target state. But, real systems interact with their environment to a greater or lesser extent. No matter how weak the coupling between the system and the environment is, the evolution of an open system will be characterised by nonunitary effects like decoherence and dissipation \cite{bib:deco}. Decoherence, in particular, is a quantum effect whereby the system loses its ability to exhibit coherent behavior. It is a serious obstacle in quantum information processing in general, and in quantum control in particular  \cite{bib:altafini}.  Control faces major difficulties when dealing with open systems, as the control field cannot fully compensate the dissipative and decohering effects from the environment. Facing this threat requires a control protocol not only to assure high fidelity but also to be implemented in a time span shorter than the decoherence time \cite{bib:sugny,bib:caneva}. Nevertheless, quantum mechanics imposes fundamental limits on the maximum rapidity with which a state can evolve to an orthogonal state. Known as the quantum speed limit (QSL), it is a physical bound that any control strategy should take into account \cite{bib:pfeifer,bib:bhatta,bib:qsl}.

In this paper we implement an efficient method for controlling a  quantum system in contact with an environment. The method is based on the knowledge of the spectrum of the closed system as a function of a suitable external parameter and requires that the system behaves locally -near avoided crossings- as the Landau-Zener (LZ) two level model \cite{bib:zener,bib:landau}. Although this characteristic may seem rather restrictive, it is, in fact, a general property of systems with interaction between its energy levels \cite{majo,bib:jopt}, at least in the low energy region.  We apply a well--defined series of fast (diabatic) and sudden (step--like) variations of the control  parameter, which allows us to travel through the state space of the system and reach the desired target state.  The speed of this control protocol meets the fundamental bounds imposed by the QSL, thus making this scheme ideal for application to open systems. We show that  this method is very successful to accomplish ambitious control goals in systems exposed to an important environmental decoherence. Additionally, this method avoids the need of using optimization algorithms in order to find a time--optimal control function, and thus various potentially encumbered subjects such as monotonicity, convergence and performance of the algorithm. 


\textit{Control strategy and QSL} - The building block of this strategy is the LZ model for a quantum system with two interacting levels. In the diabatic basis $\left\{\Ket{0},\Ket{1}\right\}$ the hamiltonian of this system can be written
\begin{equation}
 H_{LZ}=\left( 
  \begin{array}{c c}
  \alpha\lambda & \frac{\Delta}{2} \\
  \frac{\Delta}{2} & -\alpha\lambda
  \end{array} \right)= \alpha\lambda\:\sigma_z+\frac{\Delta}{2}\sigma_x,
  \label{ec:hlz}
\end{equation}
\noindent where $\Delta$ is a constant and we set $\hbar=1$ along the Letter. The eigenvalues of $H_{LZ}$ form an hyperbola in the $(\lambda,E)$ plane (as shown in Fig. 1 (a)), whose vertex represents an avoided crossing (AC) with an energy gap $\Delta$. The eigenvectors of the hamiltonian form the so-called adiabatic basis, $\left\{\Ket{\phi_0},\Ket{\phi_1}\right\}$, where the notation makes explicit the asymptotic correspondence between both basis when $\lambda\rightarrow-\infty$. When $\lambda\rightarrow+\infty$, this correspondence is exchanged. The classic LZ theory describes the transition probability of the system when the initial state is $\Ket{\psi(-\infty)}=\Ket{0}$ and the parameter $\lambda$ is sweept linearly in time, i.e. $\lambda(t)=v\:t$, yielding the famous LZ formula
\begin{equation}
P_1(t\rightarrow\infty)=1-\expo{-\frac{\pi\Delta^2}{4 v|\alpha|}}.
\label{ec:formulalz}
\end{equation}
This result defines a critical velocity $v_c=\frac{\pi\Delta^2}{4|\alpha|}$ which determines two limiting control scenarios. In the first place, we have the fast \textit{diabatic} (D) transitions, in which $v\gg v_c$ in such a way that $P_1(t\rightarrow\infty)\simeq0$, leaving the initial state unchanged. On the other hand, the \textit{adiabatic} (A) transitions, in which $v\ll v_c$ and thus $P_1(t\rightarrow\infty)\simeq1$, take place when the state evolves slowly following the adiabatic curve and finishes in state $\Ket{1}$. For a closed quantum system in which the transitions between neighbouring levels are well described by the LZ model, this scheme provides a quantitative binary (A-D) recipe for determining the appropiate velocity of the driving field at each of the avoided crossings (ACs) \cite{bib:wis1,bib:wis2,bib:wis3}. However, in the presence of an environment, decoherence is bound to act within the long periods of time required by adiabatic transitions, rendering the effective dynamics of the system non--unitary and thus preventing the system from being controlled.

To overcome this problem we propose an alternative method in which we use the so-called sudden-switch transitions \cite{bib:tambo,bib:wis2}. Consider the LZ system prepared initially in the state $\Ket{\psi(0)}=\Ket{0}$. We now consider $\lambda(t)$ to be a piecewise constant function with initial value $\lambda(0)=-\lambda_0$, with $\lambda_0\gg\Delta/\vert\alpha\vert$ \cite{majo}. In this way, the initial state is approximately an instantaneous eigenstate of the Hamiltonian $H_{LZ}(0)$. If now $\lambda(t)$ undertakes a sudden variation to $\lambda=0$ and the system is left to evolve for a time T, the final state is given by
\begin{equation}
\Ket{\psi(T)}=\mathrm{cos}\left(\frac{\Delta}{2}T\right)\Ket{0}+\mathrm{sin}\left(\frac{\Delta}{2}T\right)\Ket{1}.
\end{equation}
It is clear that choosing $T=\frac{\pi}{\Delta}$ yields $\Ket{\psi(T)}=\Ket{1}$. The final step in the evolution is a second sudden switch of the control parameter from 0 to $+\lambda_0$. Once again, the instantaneous eigenstates of the hamiltonian will be approximately those of $\sigma_z$, in such a way that $\Ket{\psi(t)}$ will be a stationary state for $t>T$, since $\Ket{\psi(T)}=\Ket{1}$. In sum, we have driven the system from $\Ket{0}$ to $\Ket{1}$ in a time $T=\frac{\pi}{\Delta}$ with a probability of 1. In this way, this scheme represents a sort of \emph{shortcut} to adiabaticity, as the final state of the system is the same as if the parameter $\lambda(t)$ had been modified adiabatically. The total evolution time has been dramatically shortened, since an adiabatic passage through the AC requires a total time much larger than a critical time of the order of $\Delta^{-2}$ (recall the critical velocity expression from the LZ formula).  Moreover, the total evolution time $T$ is precisely the shortest possible time in which this two level system can change its state, as a consequence of the time-energy uncertainty principle. This can be seen as follows. Bhattacharyya \cite{bib:bhatta} derived from the Mandelstam-Tamm inequality \cite{bib:mandelstam} the expression $\tau\geq\mathrm{arccos}(\sqrt{P_{\tau}})/\Delta H$, in which $P_{\tau}$ is the quantum non-decay probability, i.e., $P_{\tau}=\left|\BraKet{\phi}{\phi_{\tau}}\right|^2$, where $\Ket{\phi_t}=\expo{-i\frac{H}{\hbar}t}\Ket{\phi}$. In the context of control theory, $\tau$ can be chosen as the total time of the control protocol (for the two level system), in such a way that $\sqrt{P_\tau}=\left|\BraKet{\psi(0)}{\psi_{goal}}\right|$. Evaluating this result for the hamiltonian (\ref{ec:hlz}), with $\Ket{\psi(0)}=\Ket{0}$ and $\Ket{\psi_{goal}}=\Ket{1}$ we get the minimum possible $\tau$ to be $\tau_{QSL}=\pi/2\Delta H_0$. $\Delta H_0$ can be calculated straightforwardly recalling from (\ref{ec:hlz}) that $H_0=H(\lambda=0)=\frac{\Delta}{2}\sigma_x$. This yields the result $\tau_{QSL}=\frac{\pi}{\Delta}=T$, i.e., exactly the total evolution time of the sudden-switch transition, which makes it the best possible candidate for replacing slow adiabatic driving at an avoided crossing, as its time-optimality favors the prospect of avoiding decoherence.\\

\begin{figure}[h!]
\begin{center}
\includegraphics[width=\linewidth]{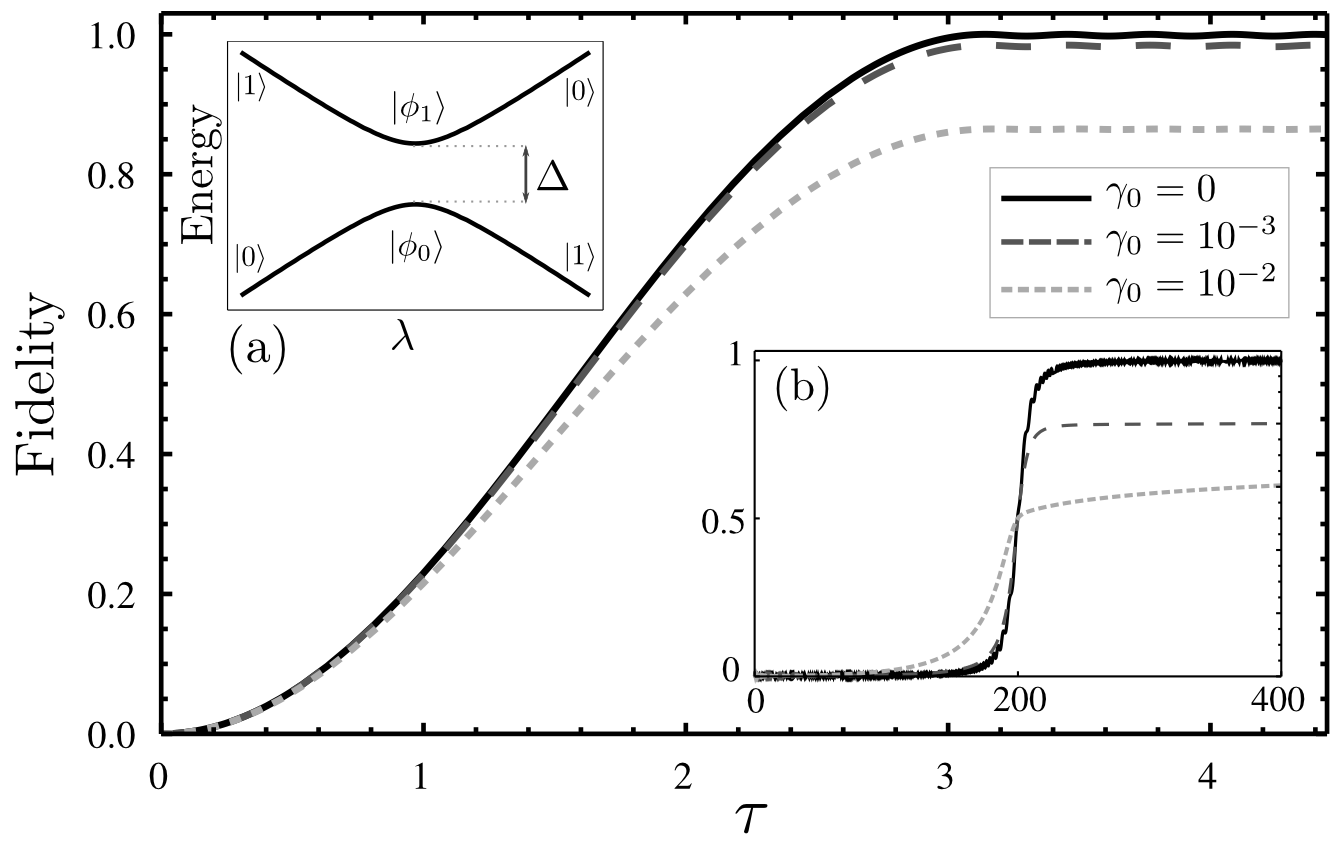}
\caption{\label{fig:fig1} \footnotesize{Fidelity as function of the dimensionless time parameter $\tau=t\Delta$ for sudden-switch variations of the control parameter. (a) Energy spectrum as a function of parameter $\lambda$ for the LZ hamiltonian (\ref{ec:hlz}). The asymptotic correspondence between the diabatic and adiabatic states has been made explicit. (b) Fidelity as a function of $\tau$ for adiabatic passages through the AC. In both fidelity plots, the solid line corresponds to the closed-system case, and the dashed curves to the system coupled to a bosonic environment. The temperature of the bath was fixed at $T=10\Delta$.}}
\end{center}
\end{figure}

\textit{Controlling the dissipative LZ model} - We now show how our control method gives satisfactory results when applied to the LZ hamiltonian coupled to a bosonic environment. We propose a master equation approach in the weak coupling and high temperature limit, which becomes independent of the particular election of time dependence of the control parameter. Assuming bilinear coupling of the form $\sigma_z\otimes\sum_j c_j q_j$ (the $q_j$'s being the position operators of the environmental oscillators), the master equation for the reduced density matrix $\rho(t)$ results \cite{bib:deco} 

\begin{equation}
\begin{split}
\dot{\rho}(t)&=-i\left(H'(t)\rho(t)-\rho(t)(H')^{\dag}(t)\right)-\frac{\gamma_0 T}{2}\conm{\sigma_z}{\conm{\sigma_z}{\rho(t)}} \\ 
& +i\frac{\gamma_0\Delta}{4}\:\left(\sigma_y\rho(t)\sigma_z-\sigma_z\rho(t)\sigma_y\right),
\end{split}
\label{ec:ecmaestra}
\end{equation}
\noindent where $\gamma_0$ is the (dimensionless) coupling constant between the system and the environment at equilibrium temperature $T$ (we set the Boltzman constant  $k_B=1$). $H'(t)=\lambda(t)\sigma_z+\frac{\Delta}{2}\left(1-i\frac{\gamma_0}{2}\right)\sigma_x$ is the renormalized non-hermitian effective hamiltonian of the system. This differential equation describes the non-unitary evolution of the reduced density matrix $\rho(t)$ of the system which, in the small coupling limit, will be mainly due to the double-commutator term. We wish to study how this non-unitary dynamic affects the fidelity of the control scheme, defined in this context as $\mathcal{F}(t)=Tr\left(\rho(t)\rho_{goal}\right)$.

It has been shown \cite{bib:lzdis1,bib:lzdis2} by different analytical approaches that while diabatic ($\rho_{goal}=\KetBra{0}{0}$) transitions in a dissipative LZ do not seem to suffer decoherence and thus achieve high fidelity, adiabatic ($\rho_{goal}=\KetBra{1}{1}$) variations of the control parameter render poor final state fidelity even for weak coupling. This can be seen in Fig. \ref{fig:fig1} inset (b), where the fidelity as a function of time is plotted for different values of $\gamma_0$. As $\lambda(t)$, being sweept slowly, reaches the AC, the system becomes extremely sensitive to decoherence and rapidly becomes mixed. The fidelity then fails to achieve the desired value of 1. For sufficiently large $\gamma_0$, the state losses its purity and evolves to $\rho=I$, for which $\mathcal{F}_{final}=\frac{1}{2}$. However, the desired transition to $\rho_{goal}=\KetBra{1}{1}$ is indeed achieved with high probability if the system is driven by the sudden switch method. In Fig. \ref{fig:fig1} we show how the fidelity in this case evolves favorably even in the presence of the environment, making this method much more robust under the action of external influences, due to its time-optimality. Moreover, the difference in the total evolution time between both methods is remarkable, as the adiabatic driving takes up to three orders of magnitude more time than the sudden-switch method. In sum, this represents a time-optimal, relatively decoherence-resistant control recipe for achieving ``adiabatic'' ($P_1\simeq1$) transitions at an AC in the energy spectrum of a system. Together with the diabatic transitions, this updated binary control protocol  (which we will refer to as S-D) renders a shorter control time and high fidelity at each AC, even in the presence of a dissipative environment.\\

\begin{figure}[h!]
\begin{center}
\includegraphics[width=\linewidth]{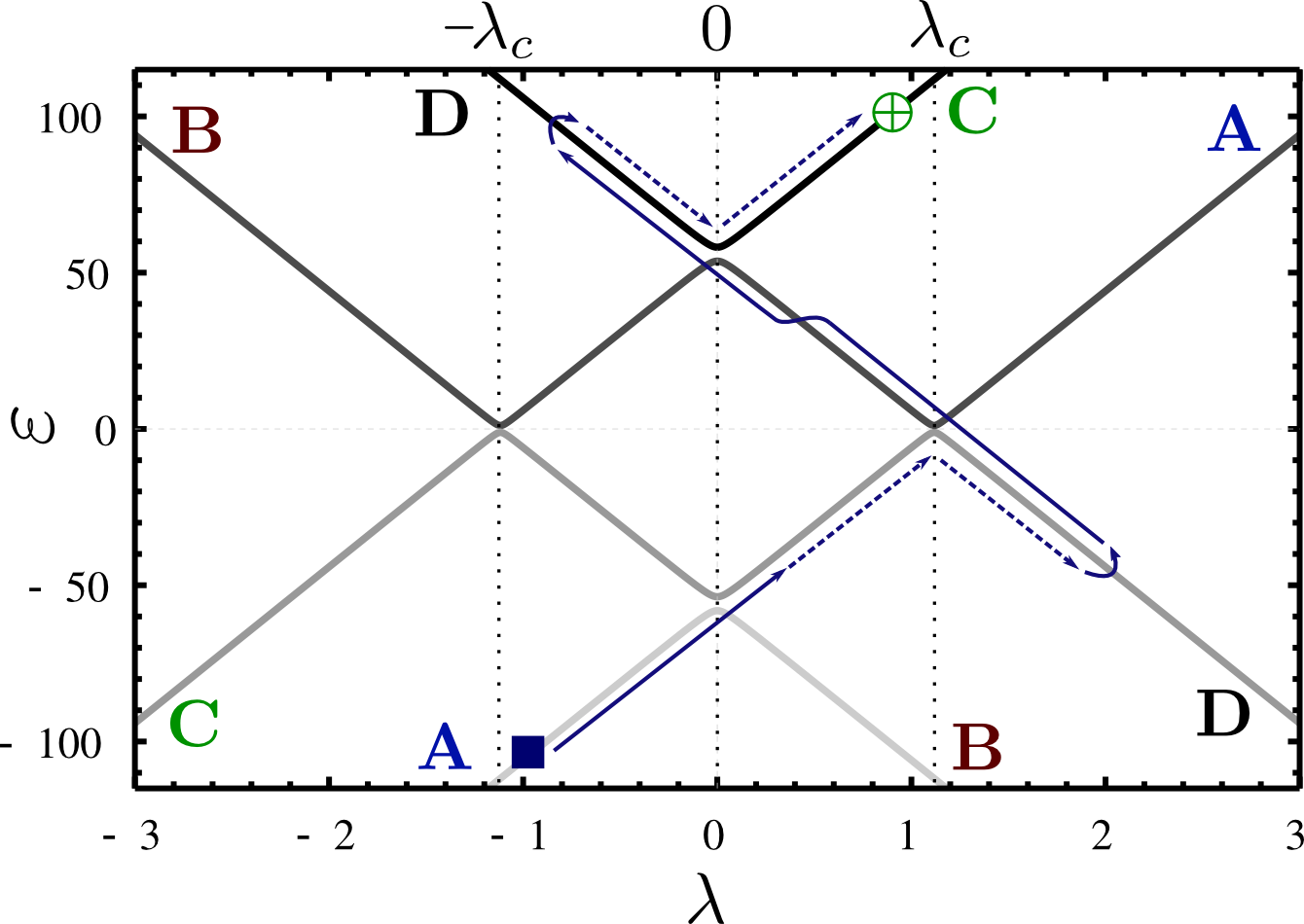}
\caption{\label{fig:fig2} \footnotesize{Spectrum of the four--level system (\ref{ec:h4niveles}), i. e., dimensionless energy $\varepsilon=\frac{E}{\Delta}$ as a function of parameter $\lambda$. The remaining parameters of the hamiltonian were set as $\Delta_A=50$ and $\Delta_B=0.05\Delta_A$. The side ACs have a gap of $\Delta=2.43$, while the central ACs have wider gaps of size $2\Delta$. The capital letters refer to the diabatic state corresponding to each energy branch at $\lambda\rightarrow\pm\infty$. The symbols $\blacksquare$ and $\oplus$ refer to the initial and target states, respectively, which are joined by a path following the energy levels. Full segments of the path symbolize diabatic variations of the control parameter, while dashed lines refer to sudden--switch transitions.}}
\end{center}
\end{figure}

\textit{Controlling a dissipative multilevel system} - We will show now how to apply our control method to a more complex multilevel system. Consider two interacting spin--$\frac{1}{2}$ particles $\mathcal{S}_\mathcal{A}$ and $\mathcal{S}_\mathcal{B}$, with dynamics given by the hamiltonian
\begin{equation}
H=\Delta_\mathcal{A}\:\sigma_x^{(\mathcal{A})}+\omega_\mathcal{B}\:\sigma_z^{(\mathcal{B})}+\Delta_\mathcal{B}\:\sigma_x^{(\mathcal{B})}+\delta\:\sigma_z^{(\mathcal{A})}\sigma_z^{(\mathcal{B})},
\label{ec:h4niveles}
\end{equation}
\noindent where $\sigma_i^{(\mathcal{X})}$ represents the i--th Pauli operator in Hilbert space associated to particle $\mathcal{X}$ and all energy parameters are fixed but $\omega_\mathcal{B}=\lambda\:\Delta_\mathcal{A}$. In Fig. \ref{fig:fig2} we show the energy spectrum associated to hamiltonian (\ref{ec:h4niveles}) as a function of $\lambda$. This spectrum resemble the one corresponding to realizable systems like a qubit in a cavity or a Josephson junction qubit \cite{exp}. The spectrum is found to have four ACs where the levels interact; two of them are located at $\lambda=0$ while the other two are placed symmetrically at $\lambda=\pm\lambda_c$. We choose the parameter $\delta=0.5\Delta_\mathcal{A}$, in such a way that two of the ACs have a gap $\Delta$ and the remaining two have a wider gap of $2\Delta$. The correspondence between the instantaneous eigenstates of the hamiltonian $H(\lambda)$ and the diabatic basis $\{\Ket{\phi_A},\Ket{\phi_B},\Ket{\phi_C},\Ket{\phi_D}\}$ has been made explicit in the figure. Since we are specially interested in studying how the fidelity of the method evolves in the presence of decoherence, we once again consider a thermal bosonic environment coupled to the subsystem $\mathcal{S}_\mathcal{A}$ bilinearly as $\sigma_z^{(\mathcal{A})}\otimes\sum_j c_j q_j$. The master equation for the reduced density matrix of the composite system, in the small coupling and high temperature limit, turns to be identical in form to (\ref{ec:ecmaestra}), where $\sigma_z$ and $\Delta$ are to be replaced with $\sigma_z^{(\mathcal{A})}$ and $\Delta_\mathcal{A}$.

\begin{figure}[h!]
\begin{center}
\includegraphics[width=0.9\linewidth]{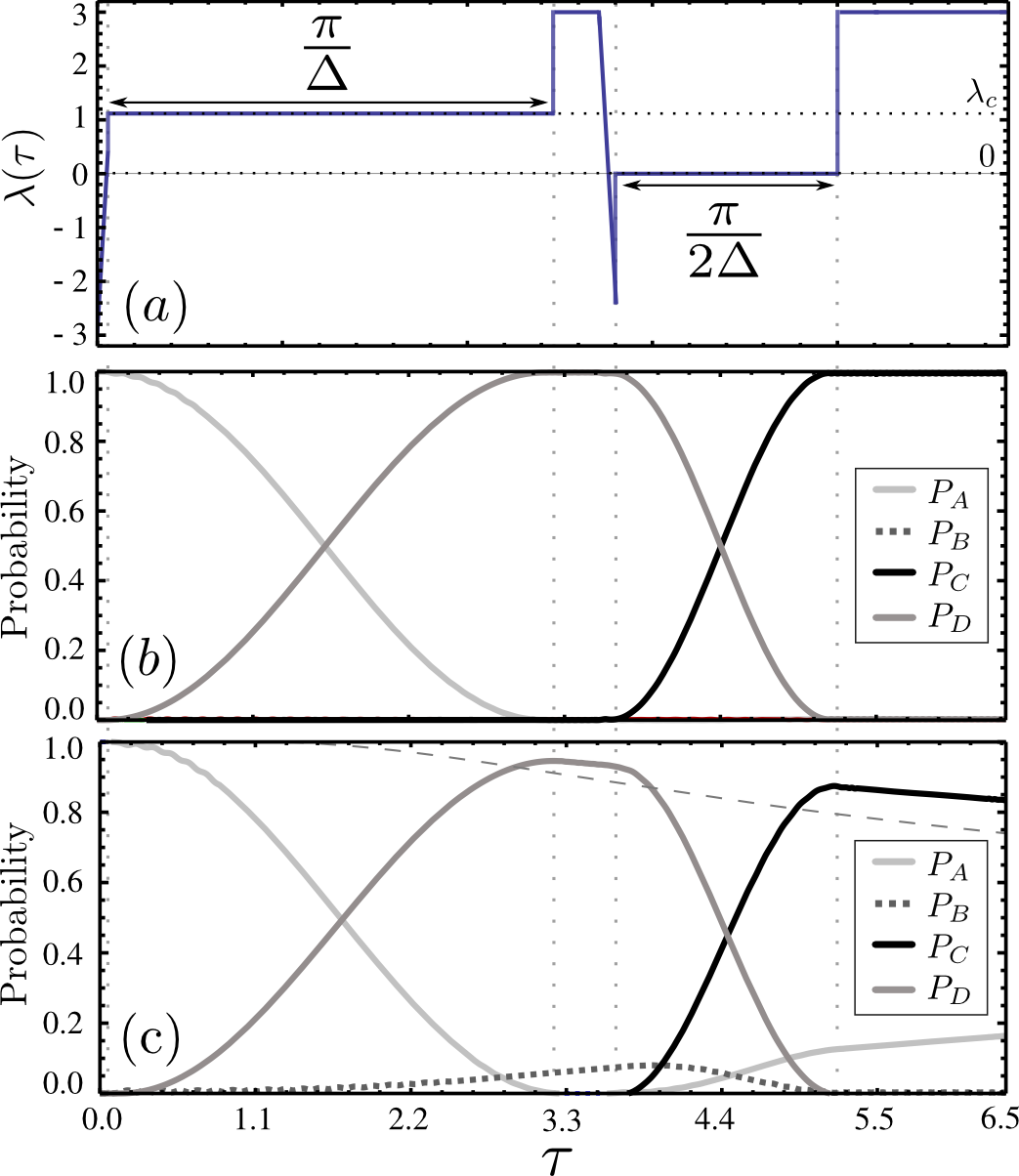}
\caption{\label{fig:fig3} \footnotesize{(a) Control parameter $\lambda$ as a function of the dimensionless time parameter $\tau=t\Delta$, designed to achieve the path displayed in the spectrum of Fig. \ref{fig:fig2} (see text for details). (b) Probabilities of finding the four-level system in each of the four diabatic states as a function of $\tau$. The dotted gray vertical lines indicate the times at which the two steps of the sudden-switches are performed. (c) Same as (b) but when the method is applied to the system coupled to an environment, with $\gamma_0=10^{-3}$. The final fidelity is $\sim0.85$. The dashed curve represents the evolution of the purity of the state. The parameters of the system are the same as those specified in Fig. \ref{fig:fig2}. The temperature was set to $T=20\Delta$.}} 
\end{center}
\end{figure}

We now pose the problem of starting from an initial state $\psi(0)=\phi_A$ and driving the system to $\psi_{goal}=\phi_C$. For that purpose, we design a path in the energy spectrum, which can be seen in Fig. \ref{fig:fig2}. This path requires the control parameter to be sweept following the sequence D-S-D-S (where D stands for a diabatic transition and S for sudden-switch transition) in order to achieve the desired navigation of the spectrum. In Fig. \ref{fig:fig3} we show the time dependence chosen for $\lambda(t)$ in order to achieve this evolution. The high--slope linear segments correspond to diabatic transitions, and two of the the constant segments have a lenght $\pi/\Delta$ and $\pi/2\Delta$, respectively, as required by the sudden-switch method.


\begin{figure}[h!]
\begin{center}
\includegraphics[width=0.9\linewidth]{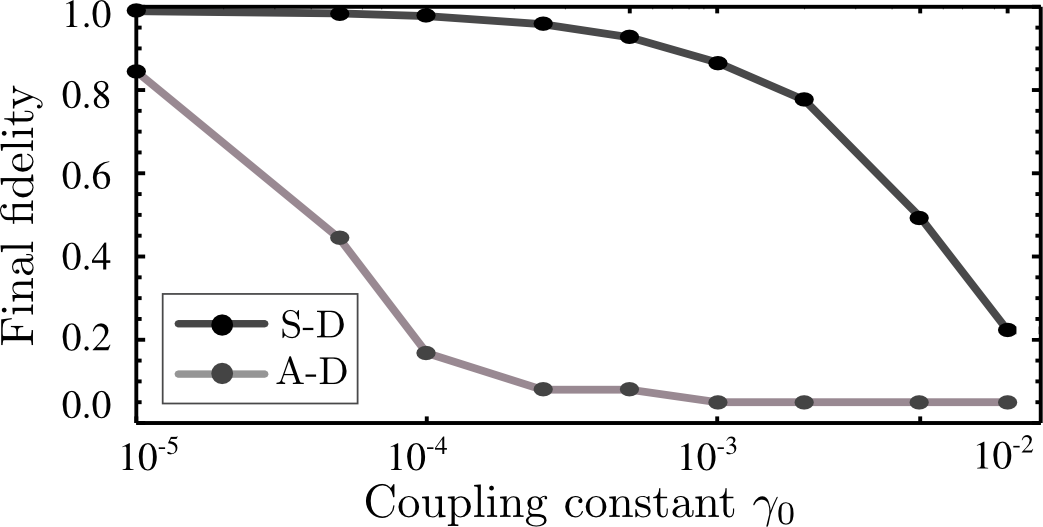}
\caption{\label{fig:fig4} \footnotesize{Final fidelity $\mathcal{F}(T)$ as a function of the dimensionless coupling constant $\gamma_0$, for the 
four-level system, using S-D and A-D control methods.}}
\end{center}
\end{figure}

In Fig. \ref{fig:fig3} (b) we show the evolution of the probabilities of finding the system in each of the various diabatic states, when the driving field of Fig. \ref{fig:fig3} (a) is applied to the closed system ($\gamma_0=0$). The fidelity in this case is $\mathcal{F}(t)=P_C(t)$, i.e., the black curve in the figure. There, it can be seen how the state remains unchanged, as expected, after the first diabatic passage. Then, as the control parameter undertakes the first sudden-switch, the state evolves to $\phi_D$ in a time $\pi/\Delta$. Next, two sucesive diabatic transitions take place when the control parameter is sweept linearly with high speed, after which the state of the system is still $\phi_D$. Finally, a second sudden switch evolves the system from $\phi_D$ to $\phi_C$, now in a time $\pi/2\Delta$. Each step of the protocol is successful, yielding an elevated final fidelity of $\mathcal{F}(T)\simeq0.99$. Moreover, the total time of evolution has barely added up to $2\pi/\Delta$, which represents an excellent time performance. The results of applying the same control protocol to the system coupled to the environment ($\gamma_0\neq0$) are shown in Fig. \ref{fig:fig3} (c). In this case, the non-unitary fashion of the evolution manifests itself through the decay of the purity of the state in time, although the rapidness of the control protocol allows it to act in a time interval shorter than the decoherence time, making the system follow the desired path and consequently achieving a fairly high fidelity. Moreover, as most of the total evolution time is due to the sudden-switch transitions, which are time-optimal, its bound to expect that there is no room for improving the final state fidelity with this or any other kind of unitary control. In Fig. \ref{fig:fig4} we show how the final fidelity $\mathcal{F}$ varies as the coupling with the environment is increased. There it can be appreciated how the S-D method yields a fidelity of over 0.8 for $\gamma_0\leq10^{-3}$, and becomes unsuccessful for sufficiently large $\gamma_0$, as can be expected due to the presence of an strongly decohering environment. For comparative purposes, a similar plot is shown with the results obtained applying the previous A-D method to the four--level system, where we can see that the old method fails even for small coupling.\\

\textit{Final Remarks} -  We have implemented an efficient method to control the state of a quantum open system. The method is based on the navigation in the energy spectrum of the closed system, using fast variations of a control parameter (which are already familiar  schemes used in NMR, quantum optics, and cavity and circuit QED, for engineering quantum dynamics). The success of this method to overcome the adverse influence of decoherence relies in the speed in which the transitions are performed. We show that the sudden transitions occur in the quantum speed limit.  The method was succesfully applied in a system with four energy levels to reach desired target states that lie far in the spectrum from the initial state. We stress that our proposed control function has a simple analytical form even for complex multilevel systems, as opposed to those obtained through optimization methods (see for example \cite{bib:natphys}). This fact makes our method very attractive for future experimental applications where several of the requirements of the method are fulfilled, but decoherence appears as an obstacle for the quantum control.

\section*{Acknowledgements}

We would like to thank Pablo Tamborenea for useful discussions. The authors acknowledge the support from CONICET, UBACyT,  and ANPCyT, Argentina.

\end{document}